\documentclass[preprint,amsmath,amssymb,showpacs,superscriptaddress,showkeys,floatfix,aps,prb]{revtex4-1}

\usepackage{graphicx}
\usepackage{amssymb}
\usepackage{amsmath}
\usepackage{dcolumn}
\usepackage{multirow}
\usepackage{bm}
\usepackage[latin1]{inputenc}
\usepackage[flushmargin,ragged]{footmisc}
\usepackage{rotating}

\begin{document}
\bibliographystyle{apsrev4-1}

\title{Group Theory analysis of phonons in two-dimensional Transition Metal Dichalcogenides}

\author{J. Ribeiro-Soares}
\email[Author to whom correspondence should be addressed: ]{jenainassoares2@gmail.com}
\affiliation{Departamento de F\'{\i}sica, Universidade Federal de Minas Gerais, Belo Horizonte, MG, 30123-970, Brazil}
\affiliation{Department of Electrical Engineering and Computer Science, Massachusetts Institute of Technology (MIT), Cambridge, MA 02139, USA}

\author{R. M. Almeida}
\affiliation{Departamento de F\'{\i}sica, Universidade Federal de Minas Gerais, Belo Horizonte, MG, 30123-970, Brazil}

\author{E. B. Barros}

\affiliation{Department of Electrical Engineering and Computer Science, Massachusetts Institute of Technology (MIT), Cambridge, MA 02139, USA}
\affiliation{Departamento de Física, Universidade Federal do Ceará, Fortaleza, CE, 60455-900, Brazil}

\author{P. T. Araujo}
\affiliation{Department of Physics and Astronomy, University of Alabama, Tuscaloosa, Alabama 35487, USA}

\author{M. S. Dresselhaus}
\affiliation{Department of Electrical Engineering and Computer Science, Massachusetts Institute of Technology (MIT), Cambridge, MA 02139, USA}
\affiliation{Department of Physics, Massachusetts Institute of Technology (MIT), Cambridge, MA 02139, USA}

\author{L. G. Can\c{c}ado}
\affiliation{Departamento de F\'{\i}sica, Universidade Federal de Minas Gerais, Belo Horizonte, MG, 30123-970, Brazil}

\author{A. Jorio}
\affiliation{Departamento de F\'{\i}sica, Universidade Federal de Minas Gerais, Belo Horizonte, MG, 30123-970, Brazil}

\date{Submitted on \today}

\begin{abstract} Transition metal dichalcogenides (TMDCs) have emerged as a new two dimensional materials field since the monolayer and few-layer limits show different properties when compared to each other and to their respective bulk materials. For example, in some cases when the bulk material is exfoliated down to a monolayer, an indirect-to-direct band gap in the visible range is observed. The number of layers $N$ ($N$ even or odd) drives changes in space group symmetry that are reflected in the optical properties. The understanding of the space group symmetry as a function of the number of layers is therefore important for the correct interpretation of the experimental data. Here we present a thorough group theory study of the symmetry aspects relevant to optical and spectroscopic analysis, for the most common polytypes of TMDCs, i.e. $2Ha$, $2Hc$ and $1T$, as a function of the number of layers. Real space symmetries, the group of the wave vectors, the relevance of inversion symmetry, irreducible representations of the vibrational modes, optical selection rules and Raman tensors are discussed.
\end{abstract}


\maketitle

\section{\label{sec:level1}Introduction}

The interest in two-dimensional layered materials was enhanced after the successful isolation of monolayer graphene (the $2$D component of graphite) reported in $2004$.\cite{novoselov2004electric} The monolayer of hexagonally linked carbon atoms made it possible to study a brand-new set of magnetic, electric and optical phenomena related to the Dirac-like nature of graphene electrons. \cite{katsnelson2006chiral} The lack of a band gap, however, imposes some difficulties to graphene's application in electronics, despite its high carrier mobility.

Other classes of $2$D materials are now also being intensively studied for many different applications motivated mainly by the need of a band gap. Perovskite-based oxides, van der Waals solids, such as Bi$_2$Se$_3$, Bi$_2$Te$_3$,\cite{zhang2009topological} hexagonal boron nitride (h-BN), \cite{kim2011synthesis} and transition metal dichalcogenides (TMDCs), such as MoS$_2$ and WSe$_2$, \cite{splendiani2010emerging,gutierrez2012extraordinary,tonndorf2013photoluminescence} offer a wide range of compounds and combinations with potential use in the emerging field of $2$D heterostructures\cite{geim2013van} (for example, tunable optoelectronic properties are obtained by a suitable choice of component layers \cite{terrones2013novel,fang2014strong}). The TMDCs are layered materials of the form MX$_2$, where ``M'' stands for groups $4-10$ of transition metals and ``X'' stands for the chalcogen atoms S, Se or Te.\cite{chhowalla2013chemistry} The ``M'' and ``X'' atoms are strongly linked through covalent bonds to form $2$D layers. Two adjacent sheets of chalcogen atoms are separated by a sheet of transition metal atoms in an X-M-X configuration, and the ``monolayer'' is actually composed of an atomic trilayer (TL) structure. The interaction among these trilayers are weak van der Waals interactions. The difference in the stacking order gives rise to different polytypes, while the combination of these different atoms leads to a variety of more than $30$ different layered materials, with different optical, mechanical and electrical properties.\cite{wang2012electronics,chhowalla2013chemistry,butler2013progress}

Some semiconducting TMDCs in this so-called monolayer form show a direct band gap in the visible range, which does not exist in their bulk counterparts.\cite{splendiani2010emerging,shaw2014chemical,gutierrez2012extraordinary,tonndorf2013photoluminescence,sahin2013anomalous} These band gaps open the possibility for flexible and transparent sensor applications,\cite{wang2012electronics,chhowalla2013chemistry,britnell2013strong} and the construction of heterostructures offers the possibility of tuning the TMDC behavior.\cite{terrones2013novel,britnell2013strong,fang2014strong} The breaking of inversion symmetry in the monolayer, with the strong spin-orbit interaction coming from the metal $d$ orbitals, gives rise to the spin splitting of the valence band at the high-symmetry $K$ points of the Brillouin Zone (BZ).\cite{xiao2012coupled} Since the $K$ and $K'$ points in the BZ are related to each other by time reversal symmetry, the spin splitting yields distinct symmetries from these two valleys, and the manipulation of this coupling opens the possibility of a variety of valleytronic applications.\cite{yao2008valley,cao2012valley,mak2012control,xiao2012coupled,zeng2013optical,xu2014spin}

The dependence on the number of layers ($N$) and on the changes of the symmetry group have already been investigated in the characterization of the various TMDC optical properties, by means of Raman spectroscopy and Second Harmonic Generation (SHG).\cite{zhao2013interlayer,luo2013effects,yamamoto2014strong,malard2013observation,zeng2013optical,li2013probing,kumar2013second,yin2014edge} Group theory provides a valuable theoretical tool that can be used to understand the selection rules for the optical transitions, to find the eigenvectors for the lattice vibrations, and to identify the lifting of degeneracies due to external symmetry-breaking perturbations.\cite{conley2013bandgap,wang2013raman} A detailed study of these symmetry aspects for few-layers TMDCs is valuable to predict interesting characteristics and to properly interpret experimental results for these compounds, since few-layers TMDCs will belong to different space groups according to the number of layers, and their space groups will be different from those of their bulk crystal counterparts.

Group theory has already been used to describe the structure of TMDCs in the bulk form, for different polytypes,\cite{wilson1969transition,katzke2004phase} in the few-TL $2Hc$ polytype for zone center phonons (at the $\Gamma$ BZ point)\cite{zhao2013interlayer,luo2013effects,yamamoto2014strong} and electronic structure at the $\Gamma$ and $K$ points,\cite{Kormanyos2013Spin_OrbitWarping} and for more detailed understanding of some non-linear optical processes.\cite{malard2013observation} In this work, group theory is applied to TMDCs in both the trigonal prismatic ($H$) and octahedral ($T$) metal atom coordinations, considering the stacking order for $2Ha$ and $2Hc$ for $H$, and $1T$ for $T$, and the dependence on the number of layers $N$ (even or odd), and considering the full set of wave vectors in the BZ, i.e., going beyond the zone center. In section \ref{sec:level1symmetryanalysis}, the symmetry analysis in real space is developed for the $2H$ (section \ref{sec:level32H}) and $1T$ (section \ref{sec:level31T}) polytypes, while the reciprocal space treatment is shown in section \ref{sec:level2GWV}. The relevance of inversion symmetry for the different TMDCs polytypes is discussed in section \ref{sec:level2InvSym}. The irreducible representations for vibrational modes for few-TL TMDCs considering the high-symmetry points and lines in the BZ are presented in section \ref{sec:level2IRvib}, and the Raman and infrared selection rules are shown in section \ref{sec:level2Ramaninfrared}, while section \ref{sec:level2tensors} gives the Raman tensors. Finally, section \ref{sec:level1summarydiscuss} summarizes the main conclusions and comments on the cases of lowering of symmetry induced by strain in MoS$_2$, by engineering heterostructures, and by breaking the out-of-plane translational symmetry in WSe$_2$.

\section{\label{sec:level1symmetryanalysis}Symmetry analysis}

    \subsection{\label{sec:level2realspace}Real space symmetry}

The family of layered TMDCs is composed of several polytypes with a different number of TLs, or different metal atom coordination that form the primitive unit cell. The main polytypes under experimental and theoretical consideration nowadays (and analyzed in the present work) are the trigonal prismatic $2H$ [$2$ TLs in a trigonal prismatic coordination ($H$) are required to form the bulk primitive unit cell] and the octahedral $1T$ [$1$TL in an octahedral coordination ($T$) is required to form the bulk primitive unit cell] (see Fig. \ref{bulkcell}). Each polytype, in turn, has a monolayer ($1$TL) as a basic $2$D building block unit. The bulk crystal is made by piling up these monolayer units, namely $1H$ (trigonal prismatic or AbA coordination, where upper cases represent chalcogen atoms and lower cases represent metal atoms) and $1T$ (octahedral or AbC coordination), as can be observed in Figs. \ref{bulkcell} (a) and (b), respectively. The blue spheres represent transition metal atoms, and the orange spheres represent the chalcogen atoms. For bulk versions of these layered materials, where the out-of-plane translational symmetry is present, the lateral view of the unit cells are highlighted with red rectangles in Figs. \ref{bulkcell} (c), (d) and (e).

There are several other polytypes for stacks of more than two TLs, and at least $11$ polytypes where identified in TMDCs.\cite{katzke2004phase} For example, the unit cell of the $3R$-MoS$_2$ (with the stacking /AbA BcB CaC/)\cite{wilson1969transition,katzke2004phase} comprises $9$ atoms in $3$TLs. The treatment of these polytypes with a high number of TLs is beyond the scope of this work but, for the $3R$ case, Table \ref{tab:Wyckoffsites} summarizes some symmetry considerations and gives representative examples.

     \subsubsection{\label{sec:level32H}$2$H polytype}

\begin{figure}[h]
        \centering
            \includegraphics[width=0.53\textwidth,natwidth=156.961,natheight=137.173]{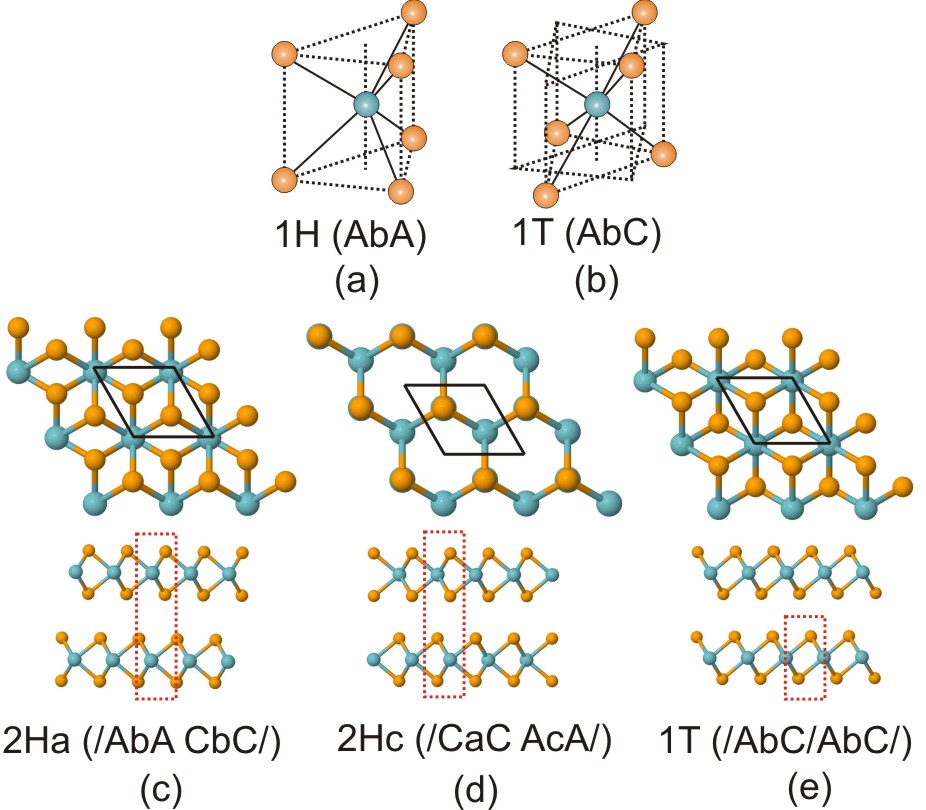}
        \caption{(Color online) Transition metal atom coordination for (a) trigonal prismatic ($H$) and (b) octahedral ($T$) TMDCs polytypes. The blue spheres represent transition metal atoms and orange ones, chalcogen atoms. In (c), (d) and (e) the top and lateral views (top and bottom in each figure, respectively) of the primitive unit cells for bulk TMDCs materials are shown. The black rhombuses show the top view of the primitive unit cell, and the red rectangles indicate the lateral view. The primitive unit cell of the $2Ha$ (c) or the $2Hc$ (d) polytypes comprise $6$ atoms, $2$ transition metal atoms and $4$ chalcogenides ($Z=2$) in the trigonal prismatic coordination illustrated in (a), while the $1T$ polytype shown in (e) has $3$ atoms, comprising $2$ chalcogenides and $1$ transition metal atom ($Z=1$) in the octahedral coordination illustrated in (b).}
        \label{bulkcell}
 \end{figure}

The $2H$ bulk polytype can assume two forms with different stacking symmetries: $2Ha$ (or /AbA CbC/ stacking),\cite{wilson1969transition,katzke2004phase} and $2Hc$ (/CaC AcA/ stacking).\cite{katzke2004phase} In $2Ha$ stacking, one transition metal atom is always on top of another transition metal atom of the next layer, as shown in Fig. \ref{bulkcell} (c). This polytype is reported to occur in NbSe$_2$, NbS$_2$, TaS$_2$ and TaSe$_2$ crystals.\cite{wilson1969transition} In $2Hc$ stacking, any transition metal atom is sitting on top of two chalcogenides atoms of the subsequent layer, as shown in Fig. \ref{bulkcell} (d). This polytype occurs in MoS$_2$, WS$_2$, MoSe$_2$ and WSe$_2$ crystals. Both polytypes belong to the non-symmorphic hexagonal space group $P6_{3}/mmc$\cite{wilson1969transition} ($D^{4}_{6h}$, in Sch\"{o}nflies notation, or \#$194$ in the International Tables for Crystallography\cite{ITCA2005}). The primitive unit cell for the bulk has $6$ atoms ($Z=2$, where $Z$ is the number of structural MX$_2$ units  required to form the primitive unit cell), and $3$ atoms in each TL, as can be seen in the red rectangles of Figs. \ref{bulkcell} (c) and (d). The Wyckoff positions for the $2H$ bulk polytypes, as well as the number of structural formulas $Z$ are given in Table \ref{tab:Wyckoffsites}.

The $2Hb$ polytype is possible and occurs for nonstoichiometric compounds with an excess of metal atoms intercalated in the van der Waals gap.\cite{katzke2004phase} Table \ref{tab:Wyckoffsites} gives symmetry information and examples for this polytype. Some differences between the definition of $2Hb$ and $2Hc$ are found in literature,\cite{wilson1969transition,katzke2004phase} and the most recent nomenclature is used in this work.\cite{katzke2004phase,hromadova2013structure}

\begin{figure}[h]
        \centering
            \includegraphics[width=0.53\textwidth,natwidth=117.247,natheight=85.329]{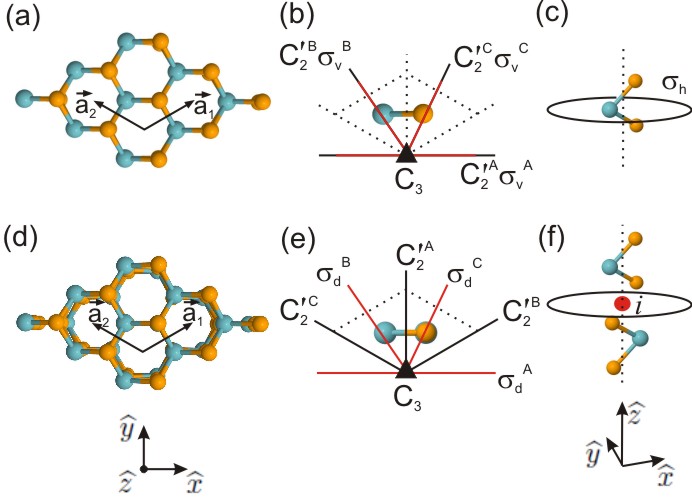}
        \caption{(Color online) Primitive unit cell and symmetry operations of the $2Hc$ polytype. Blue spheres represent transition metal atoms and orange spheres represent chalcogen atoms. (a) and (d) show the top view for the $1$TL and $2$TLs, respectively. $\vec{a}_{1}$ and $\vec{a}_{2}$ are the primitive unit vectors, indicated in (a), while (b) and (e) represent the symmetry operations for the $1$TL and $2$TLs, respectively. The $C_{3}$ axes are perpendicular to the $xy$ plane in (b) and (e), and they are represented by black triangles. Three vertical mirror planes $\sigma_{v}$ and three dihedral mirror planes $\sigma_{d}$ are indicated as red lines in (b) and (e), respectively, while the black lines are the three $C^{\prime}_2$ rotation axes in the horizontal mirror $\sigma_{h}$, represented in (c) and (f) together with the primitive unit cell. The $\sigma_{h}$ itself is not a symmetry operation for $2$TLs, but it is discussed here since it is part of the $S_{6}$ operation, which is given as a $C_{6}$ rotation followed by a $\sigma_{h}$ reflection in this plane. The red lines in (e) denote the $\sigma_{d}$ mirror planes, and the red dot in the center of (f) indicates the position of the inversion symmetry operation.}
        \label{1e2layer2Hb}
 \end{figure}

\begin{sidewaystable}
\caption{\label{tab:Wyckoffsites} Number of structural formulas ($Z$), space groups and Wyckoff positions for $2H$, $1T$ and $3R$ TMDCs polytypes. One structural formula comprises one transition metal (M) and two chalcogen atoms (X$_2$).}
\vspace{0.3cm}\tiny
\begin{ruledtabular}
\begin{tabular}{ccccccccccc}
 Polytype &\multicolumn{3}{c}{$2Ha$ polytype} &    $2Hb$ polytype\footnote{According to previous literature on TMDCs.\cite{wilson1969transition,katzke2004phase}\label{fn:repeatpreviouslit}}    &\multicolumn{3}{c}{$2Hc$ polytype} &  $3R$ polytype\footref{fn:repeatpreviouslit} & \multicolumn{2}{c}{$1T$ polytype} \\
 \hline
 Number of layers & Bulk & $N$ odd & $N$ even &   Bulk   & Bulk & $N$ odd & $N$ even  & Bulk &  Bulk & $N$ odd and $N$ even     \\
\hline
\# Structural formulas ($Z$) & $2$ & $N$ & $N$ &     $2$   & $2$ & $N$ & $N$ &  $3$ & $1$  & $N$   \\
\hline
\multirow{2}{*}{Group\footnote{The fact that $3$D space groups and the respective Wyckoff positions have been constructed considering translation along the out-of-plane direction does not change the conclusions that will be drawn in the present work because we disregard the wave vector along this non-existing direction.\label{fn:repeatwyckoff}}} & $D^{4}_{6h}$  & $D^{1}_{3h}$  & $D^{3}_{3d}$  &    $D^{1}_{3h}$    & $D^{4}_{6h}$  & $D^{1}_{3h}$  & $D^{3}_{3d}$  & $C^{5}_{3v}$ & $D^{3}_{3d}$ & $D^{3}_{3d}$    \\
 &              ($P6_{3}/mmc$, \#194) &              ($P\bar{6}m2$, \#187) &              ($P\bar{3}m1$, \#164) &       ($P\bar{6}m2$, \#187)   & ($P6_{3}/mmc$, \#194) & ($P\bar{6}m2$, \#187) & ($P\bar{3}m1$, \#164)  & ($R3m$, \#160) & ($P\bar{3}m1$, \#164) & ($P\bar{3}m1$, \#164)  \\
\hline
 \multirow{4}{*}{Wyckoff positions\footref{fn:repeatwyckoff}$^,$\footnote{The Wyckoff positions for the space groups of $N$ odd and $N$ even layers of TMDCs are not established in the International Tables of Crystallography.\cite{ITCA2005}}} &  M ($2$b) &  &  &  M$_1$ ($1$a)  & M ($2$c) &  &   & M ($3$a) & M ($1$a) &   \\
       &   &  &  &  M$_2$ ($1$d)  &  &  &  &  &  &      \\
  & X ($4$f) &  &  &  X$_1$ ($2$h)  & X ($4$f) &  &   & X$_1$ ($3$a) & X ($2$d) &     \\
       &  &  &  &  X$_2$ ($2$i)  &  &  &  &  X$_2$ ($3$a) & &   \\
\cline{2-4}
\cline{6-8}
\cline{10-11}
\multirow{2}{*}{Compounds\footref{fn:repeatpreviouslit}}  & \multicolumn{3}{c}{\multirow{2}{*}{(Nb,Ta)(S,Se,Te)$_2$}} &  Nb$_{1+x}$Se$_2$  & \multicolumn{3}{c}{Mo(S,Se,Te)$_2$}  & (Nb,Ta)(S,Se)$_2$ & \multicolumn{2}{c}{(Ti,Zr,Hf,V)(S,Se,Te)$_2$}  \\
 & \multicolumn{3}{c}{} &  Ta$_{1+x}$Se$_2$  & \multicolumn{3}{c}{W(S,Se)$_2$}   & (Mo,W)S$_2$ & \multicolumn{2}{c}{(Nb,Ta)(S,Se)$_2$}  \\
\end{tabular}
\end{ruledtabular}
\end{sidewaystable}
For few-layers systems there is a reduction in symmetry due to the lack of translational symmetry along the $z$ axis (the $z$ axis is perpendicular to the basal plane of the TLs). The symmetry operations are reduced from $24$ in the bulk to $12$ for both even and odd numbers of TLs. Therefore, the few-TLs space groups are different from the bulk space groups and depend on the parity of the number of layers (even or odd number of TLs). Figure \ref{1e2layer2Hb} illustrates $1$TL and $2$TL stacking arrangements for the $2Hc$ polytype. The hexagonal real space for $1$TL and $2$TLs are given in Figs. \ref{1e2layer2Hb} (a) and (d), respectively.

The $2Hc$ polytype symmetry operations are illustrated in Figs. \ref{1e2layer2Hb} (b) and (e), which are the top-view of the primitive unit cells. In Figs. \ref{1e2layer2Hb} (c) and (f), the lateral views of the primitive unit cells are given for $1$TL and $2$TLs, respectively. The $1$TL of $2H$ polytype belongs to the $P\bar{6}m2$ ($D^{1}_{3h}$ or \#$187$) hexagonal symmorphic space group, as well as to other few-layers compounds with odd number of layers, whose point symmetry operations are E (identity), $2C_{3}$ [clockwise and anti-clockwise rotations of $120^{\circ}$ about the axis represented as a black triangle in Fig. \ref{1e2layer2Hb} (b)], $3C^{\prime}_2$ (two-fold axis in the $\sigma_{h}$ plane), $\sigma_{h}$ (the horizontal reflection plane that passes through the transition metal atom), $2S_{3}$ ($C_{3}$ clockwise and anti-clockwise rotations, followed by a $\sigma_{h}$ reflection), and $3$$\sigma_{v}$ (vertical reflection planes).

The $2$TLs of $2H$ polytype and any other even number of TLs, belong to the $D^{3}_{3d}$ ($P\bar{3}m1$, \#164) symmorphic space group, whose symmetry operations are E, $2C_{3}$, $3C^{\prime}_2$ [rotation axes placed in between two adjacent TLs, i. e., in the middle of the van der Waals gap in Fig. \ref{1e2layer2Hb} (f)], inversion $i$ [red dot in the $\sigma_{h}$ plane of Fig. \ref{1e2layer2Hb} (f)], $3$$\sigma_{d}$ [dihedral vertical mirror planes represented by red lines in Fig. \ref{1e2layer2Hb} (e)] and $2S_{6}$ (clockwise and anti-clockwise rotations of $60^{\circ}$ followed by a $\sigma_{h}$ reflection). For the $3$TLs case, when another TL unit is added to the $2$TLs shown in Figs. \ref{1e2layer2Hb} (d), (e) and (f), the symmetry operations are the same as those observed for $1$TL, since the $\sigma_{h}$ plane is recovered as a symmetry operation. The addition of subsequent layers will always show symmetry variations depending on whether the number of layers is odd or even, and the difference between these two groups is ultimately given by the presence of the inversion symmetry in $2$TLs (which is absent in $1$TL) and the presence of the $\sigma_{h}$ plane in $1$TL (which is absent in $2$TLs).

     \subsubsection{\label{sec:level31T}$1T$ polytype}

\begin{figure}[h]
        \centering
            \includegraphics[width=0.53\textwidth,natwidth=116.646,natheight=91.804]{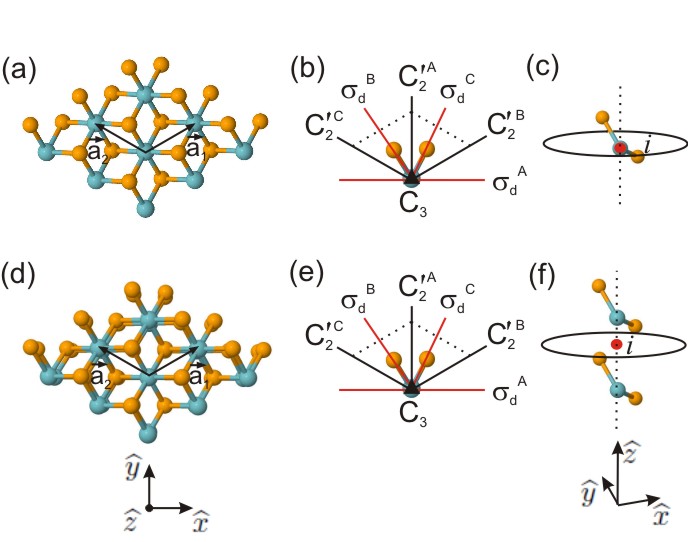}
        \caption{(Color online) Primitive unit cell and symmetry operations of the $1T$ TMDCs polytype (bulk, $1$TL and $2$TLs). (a) and (d) show the $1$TL and $2$TL top view. In (d), chalcogen atoms are on top of chalcogen atoms, and transition metal atoms are on top of transition metal atoms, giving a similar top view to that observed for $1$TL. In (b) and (e), the $C_{3}$ rotation axes (represented as black triangles) are perpendicular to the basal plane. The red lines represent $\sigma_{d}$ mirror planes, while the black lines stand for $C^{\prime}_2$ rotation axes that lie in the $\sigma_{h}$ plane. The primitive unit cells for $1$TL (and bulk) and for $2$TLs are shown in (c) and (f), respectively, and the red dot in their centers denotes the inversion operations. Notice that $\sigma_{h}$ is not a symmetry operation for $1$TL (or $N$ odd), $2$TLs (or $N$ even) or bulk, but the reflection plane is shown here to indicate the reflection in the $2S_{6}$ operations.}
        \label{1e2layer1T}
 \end{figure}

From a symmetry standpoint, the $1T$ polytype is constructed by piling up single $1$TL units, where each subsequent layer is exactly the same as the previous one, with one transition metal atom (or chalcogen atom) on top of another transition metal atom (or chalcogen atom), in an octahedral coordination. In the bulk TMDC, the stacking is /AbC/AbC/ (see Fig. \ref{bulkcell}). The bulk form belongs to the $D^{3}_{3d}$ ($P\bar{3}m1$, \#$164$) symmorphic space group. The unit cell comprises $3$ atoms of one TL [red rectangle in Fig. \ref{bulkcell} (e)]. The Wyckoff positions and number of structural formulas ($Z$) for the $1T$ polytype TMDCs are given in Table \ref{tab:Wyckoffsites}. Because all layers are identical, the symmetry operations do not change by increasing the number of TLs, no matter if $N$ is even or odd. Figures \ref{1e2layer1T} (a) and (d) show the $1$TL and $2$TLs structures, respectively, of the $1T$ polytype. The symmetry operations of $1$TL are: E, $2C_{3}$, $3C^{\prime}_2$ [the $C^{\prime}_2$ rotation axes are in the reflection plane, between the two chalcogen atoms, dividing in half the transition metal atom, as showed in the black lines in Fig. \ref{1e2layer1T} (c)], inversion $i$ (red dot in the transition metal atom), $3\sigma_{d}$ [dihedral vertical mirror planes represented by red lines in Fig. \ref{1e2layer1T} (b)] and $2S_{6}$ (clockwise and anti-clockwise rotations of $60^{\circ}$ followed by a $\sigma_{h}$ reflection). In the $2$TL case, the same operations are still valid, but now the reflection plane (Fig. \ref{1e2layer1T} (f)) for the $S_{6}$ operation is located in the van der Waals gap.

\subsection{\label{sec:level2GWV}The Group of Wave Vector}

\begin{figure}[h]
        \centering
            \includegraphics[width=0.20\textwidth,natwidth=44.006,natheight=60.266]{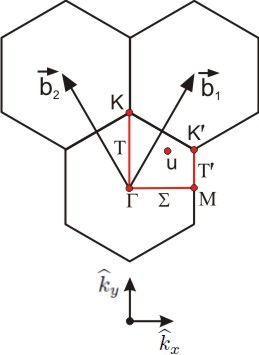}
        \caption{(Color online) The Brillouin Zone (BZ) symmetries: $\Gamma$, $K$, $K'$ and $M$ are high symmetry points; the $T$, $T'$ and
        $\Sigma$ are high symmetry lines, and the $u$ denotes the symmetry for a generic point. $\vec{b}_{1}$ and $\vec{b}_{2}$ denote the in-plane reciprocal lattice vectors.}
        \label{BZ}
 \end{figure}

The reciprocal space high symmetry points and directions for the $2H$ and $1T$ polytypes are shown in Fig. \ref{BZ}. Here $\vec{a}_{1}$ and $\vec{a}_{2}$ are the primitive vectors of the real $2$D lattice described by Eq. (\ref{a1a2real}) and are shown in Fig. \ref{1e2layer2Hb} (a). Correspondingly, $\vec{b}_{1}$ and $\vec{b}_{2}$ [described in Eq. (\ref{b1b2reciproco})] are the reciprocal lattice vectors shown in Fig. \ref{BZ}.

 \begin{align}\label{a1a2real}
\vec{a}_{1}=\,\frac{a}{2}(\sqrt{3}~\widehat{x}+\widehat{y}) && \vec{a}_{2}=\,\frac{a}{2}(-\sqrt{3}~\widehat{x}+\widehat{y})
\end{align}

 \begin{align}\label{b1b2reciproco}
\vec{b}_{1}=\,\frac{2\pi}{a}(\frac{\sqrt{3}}{3}~\widehat{k}_{x}+\widehat{k}_{y}) && \vec{b}_{2}=\,\frac{2\pi}{a}(-\frac{\sqrt{3}}{3}~\widehat{k}_{x}+\widehat{k}_{y})
\end{align}.

The differences between the space groups $D^{1}_{3h}$ and $D^{3}_{3d}$ when the number of TLs is odd or even define different symmetries for the Group of the Wave Vectors (GWV) at each high-symmetry point or direction of the reciprocal space. Knowledge of the GWV is important because the invariance of the Hamiltonian under symmetry operations usually leads to degeneracies at these high-symmetry points or directions in the BZ.\cite{tinkham2012group,dresselhaus2008group,jacobs2005group} The GWV for the $2H$ TMDCs is similar to the GWV found for $N$-layer graphene and bulk graphite,\cite{malard2009group} since the space groups for bulk, $N$ even, and $N$ odd ($N$$\geq3$) TLs in the TMDC family resemble the corresponding graphene systems. However, the $1$TL case in TMDCs lacks the inversion symmetry and therefore belongs to the same space group ($P\bar{6}m2$) as that for other $N$-odd thin layers. Table \ref{tab:GWV2H} shows the point groups that are isomorphic to the GWV for all the BZ high-symmetry points and axes occuring for bulk and for both odd or even number of TLs in $2H$ polytype.

\begin{sidewaystable}
\caption{\label{tab:GWV2H} Space groups and group of the wave vector (GWV) according to the number $N$ of TLs for all high symmetry points and lines in the BZ of the $2H$ polytype of TMDCs.}
\vspace{0.3cm}\tiny
\begin{ruledtabular}
\begin{tabular}{cccccccc}
   & Space group & $\Gamma$ & $K (K')$ & $M$ & $T (T')$ & $\Sigma$ & $u$ \\
\hline
$N$ odd & $D^{1}_{3h}$ ($P\bar{6}m2$, \#187)  & $D^{1}_{3h}$ ($P\bar{6}m2$, \#187) & $C^{1}_{3h}$ ($P\bar{6}$, \#174) & $C^{14}_{2v}$ ($Amm2$, \#38) & $C^{xy}_{s}$ (or $C^{1}_{s}$, $Pm$, \#6)\footnote{``$xy$'' is the $\sigma$'s mirror plane.\label{fn:repeatxy}} & $C^{14}_{2v}$ ($Amm2$, \#38) & $C^{xy}_{s}$ (or $C^{1}_{s}$, $Pm$, \#6) \\
$N$ even & $D^{3}_{3d}$ ($P\bar{3}m1$, \#164)  & $D^{3}_{3d}$ ($P\bar{3}m1$, \#164) & $D^{2}_{3}$ ($P321$, \#150) & $C^{3}_{2h}$ ($C2/m$, \#12) & $C^{3}_{2}$ ($C2$, \#5) & $C^{xz}_{s}$ (or $C^{3}_{s}$, $Cm$, \#8)\footnote{``$xz$'' is the $\sigma$'s mirror plane.} & $C^{1}_{1}$ ($P1$, \#1)\\
Bulk & $D^{4}_{6h}$ ($P6_{3}/mmc$, \#194)  & $D^{4}_{6h}$ ($P6_{3}/mmc$, \#194) & $D^{4}_{3h}$ ($P\bar{6}2c$, \#190) & $D^{17}_{2h}$ ($Cmcm$, \#63) & $C^{16}_{2v}$ ($Ama2$, \#40) & $C^{14}_{2v}$ ($Amm2$, \#38) & $C^{xy}_{s}$ (or $C^{1}_{s}$, $Pm$, \#6)\footref{fn:repeatxy}\\
\end{tabular}
\end{ruledtabular}
\end{sidewaystable}

   \begin{table}[h!]
\caption{\label{tab:1TGWV}Space group and group of the wave vector (GWV) for the high symmetry points and directions in the BZ for $1T$ polytype in TMDCs, valid for $N$-layer (even or odd) and bulk.}
\vspace{0.3cm}\tiny
\begin{ruledtabular}
\begin{tabular}{ccccccc}
   Space group & $\Gamma$ & $K (K')$ & $M$ & $T (T')$ & $\Sigma$ & $u$\\
\hline
  $D^{3}_{3d}$ ($P\bar{3}m1$, \#164)  & $D^{3}_{3d}$ ($P\bar{3}m1$, \#164) & $D^{2}_{3}$ ($P321$, \#150) & $C^{3}_{2h}$ ($C2/m$, \#12) & $C^{3}_{2}$ ($C2$, \#5) & $C^{xz}_{s}$ (or $C^{3}_{s}$, $Cm$, \#8)\footnote{``$xz$'' is the $\sigma$'s mirror plane.} & $C^{1}_{1}$ ($P1$, \#1)\\
\end{tabular}
\end{ruledtabular}
    \end{table}

The $1T$ polytype has the same GWV regardless of the number of layers in the sample. The bulk is symmorphic, so it has the same GWV. Table \ref{tab:1TGWV} shows the GWV for different high-symmetry points and axes within the BZ for this polytype.

\subsection{\label{sec:level2InvSym}The relevance of inversion symmetry}

The presence or absence of inversion symmetry is an important aspect of TMDCs, since it opens the possibility of coupled spin and valley physics.\cite{xiao2012coupled} The strong Spin-Orbit Coupling in TMDC materials is due to the $d$ orbitals in their heavy metal atoms. The absence of inversion symmetry lifts the degeneracy of the same energy at the same $\vec{k}$ value, at the $K$ point of the BZ, and spin splitting values on the order of $0.4$ eV have been observed in WSe$_{2}$.\cite{zeng2013optical}

The inversion symmetry is also important for the Second-Harmonic Generation (SHG) technique, which has been routinely used to probe not only the presence of inversion symmetry, but also the crystal orientation\cite{malard2013observation,li2013probing} and, recently, the effect in SHG of two artificially stacked TMDCs layers.\cite{hsu2014second} For centrosymmetric crystals, the $\chi^{(2)}$ nonlinear susceptibility vanishes,\cite{boyd2008nonlinear} and SHG signal is not observed. The $2H$ TMDCs polytype (and in this case, it also includes the $1$TL), belong to the non-centrosymmetric space group $D_{3h}^1$ and then it is possible to observe the SHG.\cite{shen1984principles,boyd2008nonlinear,malard2013observation,kumar2013second,zeng2013optical,li2013probing,hsu2014second} The $N$-even TLs for $2H$ TMDCs do not show SHG, since their space groups are centrosymmetric. For the $1T$ TMDCs polytype, both $N$-even and $N$-odd TLs have the same centrosymmetric space group $D_{3d}^3$, and the SHG signal is not expected. In this sense, the SHG mapping (together with other characterization tools) could be used to detect different polytypes in the same sample, since the $2H$ polytype with an odd number of layers shows SHG, while the layered $1T$ polytype does not.

  \subsection{\label{sec:level2IRvib}Irreducible representations for vibrational modes}

The irreducible representations for the lattice vibrations ($\Gamma^{vib}$) are given by the direct product $\Gamma^{vib}$$=$$\Gamma^{eq}$$\oplus$$\Gamma^{vec}$, where $\Gamma^{eq}$ denotes the equivalence representation for the atomic sites, and $\Gamma^{vec}$ is the representation for the $x$, $y$ and $z$ real space vectors.\cite{dresselhaus2008group} The $\Gamma^{vec}$ representation can be written as $\Gamma^{vec}$$=$$\Gamma^{x}$$\oplus$$\Gamma^{y}$$\oplus$$\Gamma^{z}$, or $\Gamma^{vec}$$=$$\Gamma^{x,y}$$\oplus$$\Gamma^{z}$ when $x$ and $y$ have the same irreducible representation. The $\Gamma^{vib}$ representations for the $2Ha$, $2Hc$ and $1T$ polytypes are given in Tables \ref{tab:Gamma2Ha}, \ref{tab:Gamma2Hb} and \ref{tab:Gamma1T}, respectively, for all the BZ high-symmetry points and lines (shown in Fig. \ref{BZ}), and for odd or even numbers of TLs. It is worth noticing that for the $2Hc$ polytype, the $\Gamma^{vib}$ for the $K'$ point is the complex conjugated form of the $\Gamma^{vib}$ for the $K$ point, while for the $2Ha$ polytype, the atomic sites are different (due to different Wyckoff positions) and the $\Gamma^{vib}$ of the $K$ and $K'$ points are the same. In the $1T$ polytype, the $\Gamma^{vib}$ for the $K$ and $K'$ points are also the same. The conversion from the Space Group (SG) to the Point Group (PG) notation for the irreducible representations is indicated in each character table of the Supplementary Material.\cite{[{See Supplementary Material at }][{ for character tables (with the notation conversion from space group to point group, for all the GWV used in this work) and for tables for the irreducible representations for lattice vibrations ($\Gamma^{vib}$) for bulk $2H$ and $1T$ polytypes.}]SupInfolink} The irreducible representations for vibrations for each high-symmetry point and line of the BZ for all the bulk polytypes are also given in Tables SI, SII and SIII of the Supplementary Material.\cite{[{See Supplementary Material at }][{ for character tables (with the notation conversion from space group to point group, for all the GWV used in this work) and for tables for the irreducible representations for lattice vibrations ($\Gamma^{vib}$) for bulk $2H$ and $1T$ polytypes.}]SupInfolink}

\begin{table*}[h!]
\caption{\label{tab:Gamma2Ha}Normal vibrational mode irreducible representations ($\Gamma^{vib}$) for $N$-layer TMDCs $2Ha$-polytype (/AbA CbC/), considering all the high-symmetry points and lines in the BZ.}
\vspace{0.3cm}\tiny
\begin{ruledtabular}
\begin{tabular}{ccc}
 \multicolumn{3}{c}{$2Ha$-polytype (/AbA CbC/)}\\ \hline
 &$N$ odd
&$N$ even\\ \hline
 $\Gamma$&$(\frac{3N-1}{2})(\Gamma_{1}^+\oplus\Gamma_{3}^-)\oplus(\frac{3N+1}{2})(\Gamma_{3}^+\oplus\Gamma_{2}^-)$&$(\frac{3N}{2})(\Gamma_{1}^+\oplus\Gamma_{3}^+\oplus\Gamma_{2}^-\oplus\Gamma_{3}^-)$ \\
 $K$($K'$)&$(\frac{3N-1}{2})(K_{1}^+\oplus K_{2}^-\oplus K^{-*}_2)\oplus(\frac{3N+1}{2})(K_{2}^+\oplus K^{+*}_2\oplus K_{1}^-)$&$(\frac{3N}{2})(K_{1}\oplus K_{2})\oplus3NK_{3}$ \\
 $M$ &$3N(M_{1}\oplus M_{4})\oplus(\frac{3N-1}{2})M_{2}\oplus(\frac{3N+1}{2})M_{3}$&$3N(M_{1}^+\oplus M_{2}^-)\oplus(\frac{3N}{2})(M_{2}^+\oplus M_{1}^-)$  \\
 $\Sigma$&$3N(\Sigma_{1}\oplus\Sigma_{4})\oplus(\frac{3N-1}{2})\Sigma_{2}\oplus(\frac{3N+1}{2})\Sigma_{3}$&$6N\Sigma_{1}\oplus3N\Sigma_{2}$ \\
 $T (T')$&$(\frac{9N+1}{2})T^+\oplus(\frac{9N-1}{2})T^-$&$(\frac{9N}{2})(T_{1}\oplus T_{2})$ \\
 $u$ &$(\frac{9N+1}{2})u^+\oplus(\frac{9N-1}{2})u^-$&$9Nu$ \\
\end{tabular}
\end{ruledtabular}
\end{table*}

\begin{table*}[h!]
\caption{\label{tab:Gamma2Hb}Normal vibrational mode irreducible representations ($\Gamma^{vib}$) for the $N$-layer TMDCs $2Hc$-polytype (/CaC AcA/), considering all the high-symmetry points and lines in the BZ.}
\vspace{0.3cm}\tiny
\begin{ruledtabular}
\begin{tabular}{ccc}
 \multicolumn{3}{c}{$2Hc$-polytype (/CaC AcA/)}\\ \hline
 &$N$ odd
& $N$ even\\ \hline
 $\Gamma$&$(\frac{3N-1}{2})(\Gamma_{1}^+\oplus\Gamma_{3}^-)\oplus(\frac{3N+1}{2})(\Gamma_{3}^+\oplus\Gamma_{2}^-)$&$(\frac{3N}{2})(\Gamma_{1}^+\oplus\Gamma_{3}^+\oplus\Gamma_{2}^-\oplus\Gamma_{3}^-)$ \\
 $K$($K'^{*}$)&$(\frac{3N+1}{2})(K_{1}^+\oplus K^{+}_2\oplus K^{-*}_2)\oplus(\frac{3N-1}{2})(K_{1}^-\oplus K_{2}^-\oplus K^{+*}_2)$&$(\frac{3N}{2})(K_{1}\oplus K_{2})\oplus3NK_{3}$ \\
 $M$ &$3N(M_{1}\oplus M_{4})\oplus(\frac{3N-1}{2})M_{2}\oplus(\frac{3N+1}{2})M_{3}$&$3N(M_{1}^+\oplus M_{2}^-)\oplus(\frac{3N}{2})(M_{2}^+\oplus M_{1}^-)$  \\
 $\Sigma$&$3N(\Sigma_{1}\oplus\Sigma_{4})\oplus(\frac{3N-1}{2})\Sigma_{2}\oplus(\frac{3N+1}{2})\Sigma_{3}$&$6N\Sigma_{1}\oplus3N\Sigma_{2}$ \\
 $T (T')$&$(\frac{9N+1}{2})T^+\oplus(\frac{9N-1}{2})T^-$&$(\frac{9N}{2})(T_{1}\oplus T_{2})$ \\
 $u$&$(\frac{9N+1}{2})u^+\oplus(\frac{9N-1}{2})u^-$&$9Nu$ \\
\end{tabular}
\end{ruledtabular}
\end{table*}

\begin{table*}[h!]
\caption{\label{tab:Gamma1T}Normal vibrational mode irreducible representations ($\Gamma^{vib}$) for the $N$-layer TMDCs $1T$-polytype (/AbC/AbC/), considering all the high-symmetry points and lines in the BZ.}
\vspace{0.3cm}\tiny
\begin{ruledtabular}
\begin{tabular}{ccc}
 \multicolumn{3}{c}{   $1T$-polytype (/AbC/AbC/)}\\ \hline
 &$N$ odd
&$N$ even\\ \hline
 $\Gamma$&$(\frac{3N-1}{2})(\Gamma_{1}^+\oplus\Gamma_{3}^+)\oplus(\frac{3N+1}{2})(\Gamma_{2}^-\oplus\Gamma_{3}^-)$&$(\frac{3N}{2})(\Gamma_{1}^+\oplus\Gamma_{3}^+\oplus\Gamma_{2}^-\oplus\Gamma_{3}^-)$ \\
 $K$($K'$)&$(\frac{3N-1}{2})K_{1}\oplus(\frac{3N+1}{2})K_{2}\oplus3NK_{3}$&$(\frac{3N}{2})(K_{1}\oplus K_{2})\oplus3NK_{3}$ \\
 $M$&$(3N-1)(M_{1}^+\oplus M_{1}^-)\oplus(\frac{3N-1}{2})M_{2}^+\oplus(3N+1)M_{2}^-$&$3N(M_{1}^+\oplus M_{2}^-)\oplus(\frac{3N}{2})(M_{2}^+\oplus M_{1}^-)$  \\
 $\Sigma$&$6N\Sigma_{1}\oplus3N\Sigma_{2}$&$6N\Sigma_{1}\oplus3N\Sigma_{2}$ \\
 $T (T')$&$(\frac{9N-1}{2})T_{1}\oplus(\frac{9N+1}{2})T_{2}$&$(\frac{9N}{2})(T_{1}\oplus T_{2})$ \\
 $u$&$9Nu$&$9Nu$ \\
\end{tabular}
\end{ruledtabular}
\end{table*}

  \subsection{\label{sec:level2Ramaninfrared}Raman and infrared selection rules}

For bulk $2H$ polytypes ($1T$ polytype), the lattice vibration irreducible representations $\Gamma^{vib}$ for the $18$ ($9$) zone center phonons are reproduced in the first line of Table \ref{tab:RamanIRACSilbulk} (see also Tables SI and SII from Supplementary Material).\cite{[{See Supplementary Material at }][{ for character tables (with the notation conversion from space group to point group, for all the GWV used in this work) and for tables for the irreducible representations for lattice vibrations ($\Gamma^{vib}$) for bulk $2H$ and $1T$ polytypes.}]SupInfolink} The classification of the modes as Raman active, infrared (IR) active, acoustic, and silent are given in Table \ref{tab:RamanIRACSilbulk}.

\begin{table*}[h!]
\caption{\label{tab:RamanIRACSilbulk} Normal vibrational mode irreducible representations ($\Gamma^{vib}$) for bulk TMDCs at the $\Gamma$ point within the $2Ha$, $2Hc$ and $1T$ polytypes. The Raman active, infrared active, acoustic and silent mode irreducible representations are identified.}
\vspace{0.3cm}\tiny
\begin{ruledtabular}
\begin{tabular}{ccc}
 &\multicolumn{1}{c}{$2Ha$ and $2Hc$ polytypes}& $1T$ polytype\\
\cline{2-2}
$\Gamma^{vib}$ & $\Gamma_{1}^+\oplus2\Gamma_{3}^+\oplus\Gamma_{5}^+\oplus2\Gamma_{6}^+\oplus2\Gamma_{2}^-\oplus\Gamma_{4}^-\oplus2\Gamma_{5}^-\oplus\Gamma_{6}^-$ & ~~~~$\Gamma_{1}^+\oplus\Gamma_{3}^+\oplus2\Gamma_{2}^-\oplus2\Gamma_{3}^-$  \\
\hline
Raman & $\Gamma_{1}^+\oplus\Gamma_{5}^+\oplus2\Gamma_{6}^+$ & $\Gamma_{1}^+\oplus\Gamma_{3}^+$  \\
Infrared & $\Gamma_{2}^-\oplus\Gamma_{5}^-$ & $\Gamma_{2}^-\oplus\Gamma_{3}^-$ \\
Acoustic & $\Gamma_{2}^-\oplus\Gamma_{5}^-$ & $\Gamma_{2}^-\oplus\Gamma_{3}^-$ \\
Silent & $2\Gamma_{3}^+\oplus\Gamma_{4}^-\oplus1\Gamma_{6}^-$ & -  \\
\end{tabular}
\end{ruledtabular}
\end{table*}

For the $2$D polytypes, the Raman and IR active modes show symmetry variations depending on the number of layers, since the high-symmetry $\Gamma$ points have different GWV. The GWV at the $\Gamma$ point is $D^{1}_{3h}$ for $N$-odd $2H$ polytypes, $D^{3}_{3d}$ for $N$-even $2H$ polytypes, and $D^{3}_{3d}$ for the $N$-even and $N$-odd $1T$ polytype. The total number of modes for $N$ even or $N$ odd layers in the $2H$ and $1T$ polytypes, including their classification as Raman active, IR active, acoustic, and silent modes, are given in Tables \ref{tab:RamanIRACSilNlayer2H} and \ref{tab:RamanIRACSilNlayer1T}, respectively.

\begin{table*}[h!]
\caption{\label{tab:RamanIRACSilNlayer2H} Normal vibrational mode irreducible representations ($\Gamma^{vib}$) for the $N$-layer TMDCs at the $\Gamma$ point within the $2Ha$ and $2Hc$ polytypes. Raman active, infrared active, acoustic and silent mode irreducible representations are identified.}
\vspace{0.3cm}\tiny
\begin{ruledtabular}
\begin{tabular}{ccc}
\multicolumn{3}{c}{{$2Ha$ and $2Hc$ polytypes}}\\
\hline
 &$N$ odd & $N$ even  \\
\hline
$\Gamma^{vib}$ & $(\frac{3N-1}{2})(\Gamma_{1}^+\oplus\Gamma_{3}^-)\oplus(\frac{3N+1}{2})(\Gamma_{3}^+\oplus\Gamma_{2}^-)$ & $(\frac{3N}{2})(\Gamma_{1}^+\oplus\Gamma_{3}^+\oplus\Gamma_{2}^-\oplus\Gamma_{3}^-)$  \\
\hline
Raman & $\frac{(3N-1)}{2}(\Gamma_{1}^+\oplus\Gamma_{3}^-\oplus\Gamma_{3}^+)$ & $\frac{3N}{2}(\Gamma_{1}^+\oplus\Gamma_{3}^+)$ \\
Infrared & $\frac{(3N-1)}{2}(\Gamma_{3}^+\oplus\Gamma_{2}^-)$ & $\frac{(3N-2)}{2}(\Gamma_{2}^-\oplus\Gamma_{3}^-)$ \\
Acoustic & $\Gamma_{3}^+\oplus\Gamma_{2}^-$ & $\Gamma_{2}^-\oplus\Gamma_{3}^-$ \\
Silent & - & - \\
\end{tabular}
\end{ruledtabular}
\end{table*}

\begin{table*}[h!]
\caption{\label{tab:RamanIRACSilNlayer1T} Normal vibrational mode irreducible representations ($\Gamma^{vib}$) for the $N$-layer TMDCs at the $\Gamma$ point within the $1T$-polytype. Raman active, infrared active, acoustic and silent mode irreducible representations are identified.}
\vspace{0.3cm}\tiny
\begin{ruledtabular}
\begin{tabular}{ccc}
 \multicolumn{3}{c}{$1T$ polytype}\\
\hline
 & $N$ odd & $N$ even \\
\hline
$\Gamma^{vib}$ & $(\frac{3N-1}{2})(\Gamma_{1}^+\oplus\Gamma_{3}^+)\oplus(\frac{3N+1}{2})(\Gamma_{2}^-\oplus\Gamma_{3}^-)$ & $(\frac{3N}{2})(\Gamma_{1}^+\oplus\Gamma_{3}^+\oplus\Gamma_{2}^-\oplus\Gamma_{3}^-)$ \\
\hline
Raman & $\frac{(3N-1)}{2}(\Gamma_{1}^+\oplus\Gamma_{3}^+)$ & $\frac{3N}{2}(\Gamma_{1}^+\oplus\Gamma_{3}^+)$ \\
Infrared & $\frac{(3N-1)}{2}(\Gamma_{2}^-\oplus\Gamma_{3}^-)$ & $\frac{(3N-2)}{2}(\Gamma_{2}^-\oplus\Gamma_{3}^-)$ \\
Acoustic & $\Gamma_{2}^-\oplus\Gamma_{3}^-$ & $\Gamma_{2}^-\oplus\Gamma_{3}^-$ \\
Silent & - & - \\
\end{tabular}
\end{ruledtabular}
\end{table*}

 In the $1T$ polytype, since the space group is the same in both $N$-even and $N$-odd, the representations for the few-TL films of this polytype refer to the same irreducible representations of the group of the wave vector $D^{3}_{3d}$ at the $\Gamma$ point, which, in turn, are the same as those found for its bulk counterpart.

\subsection{\label{sec:level2tensors}Raman tensors}

To define whether or not a specific vibrational mode will be experimentally observed in a given Raman scattering geometry, we use here the Porto notation,\cite{portonotation,damen1966raman} which indicates the crystal orientation with respect to the polarization and propagation directions of the laser. Four letters are used in the Porto notation to describe the scattering process in the a(bc)d form: while ``a'' and ``d'' are the propagation directions of the incident and scattered light, respectively, ``b'' and ``c'' represent the polarization directions for the incident and scattered light, respectively. One common Raman experimental geometry is the backscattering configuration, where the incident and scattered light have an opposite sense. For example, in the $\overline{z}(xy)z$ configuration the $\overline{z}$ and $z$ are the directions of the incident and scattered light, with the opposite sense, $x$ is the polarization direction of the incident light, and $y$ is the polarization direction of the scattered light.

The Raman scattering intensity given by the Hamiltonian perturbation term is proportional to $\mathopen|\widehat{e}_{s}.\overleftrightarrow{\alpha}\widehat{e}_{i}\mathopen|^2$, where $\widehat{e}_{s}$ is the unit vector along the polarization direction of the scattered light, $\widehat{e}_{i}$ is the unit vector along the polarization direction of the incident light, and $\overleftrightarrow{\alpha}$ is the Raman tensor. The quadratic functions ($xx$, $xy$, $xz$, $yz$...) indicate the irreducible representations for the Raman-active modes. Following this procedure, the Raman tensors for all the Raman active modes of $N$-layer thin films can be found. For the $2H$ polytype with $N$-odd few layers ($D^{1}_{3h}$ group of the wave vector for the $\Gamma$ point), the Raman tensors are:\cite{bilbaoramantensors}

$$
\Gamma_{1}^+(A'_{1}):\begin{pmatrix}
   a & 0 & 0 \\
   0 & a & 0 \\
   0 & 0 & b \\
\end{pmatrix},
$$
$$
\Gamma_{3}^+(E')_{(x)}:\begin{pmatrix}
   d & 0 & 0 \\
   0 & -d & 0 \\
   0 & 0 & 0 \\
\end{pmatrix},
\quad
\Gamma_{3}^+(E')_{(y)}:\begin{pmatrix}
   0 & -d & 0 \\
   -d & 0 & 0 \\
   0 & 0 & 0 \\
\end{pmatrix},
$$

$$
\Gamma_{3}^-(E''):\begin{pmatrix}
   0 & 0 & -c \\
   0 & 0 & 0 \\
   -c & 0 & 0 \\
\end{pmatrix},
\quad
\begin{pmatrix}
   0 & 0 & 0 \\
   0 & 0 & c \\
   0 & c & 0 \\
\end{pmatrix}.
$$

For the $N$-even $2H$ polytype, and for the $N$ even or odd for the $1T$ polytype, as well as for the $1T$ bulk crystal ($D^{3}_{3d}$ group of the wave vector for the $\Gamma$ point), the Raman tensors are:\cite{bilbaoramantensors}

$$
\Gamma_{1}^+(A_{1g}):\begin{pmatrix}
   a & 0 & 0 \\
   0 & a & 0 \\
   0 & 0 & b \\
\end{pmatrix},
$$
$$
\Gamma_{3}^+(E_{g})_{(1)}:\begin{pmatrix}
   c & 0 & 0 \\
   0 & -c & d \\
   0 & d & 0 \\
\end{pmatrix}
\quad
  \Gamma_{3}^+(E_{g})_{(2)}:\begin{pmatrix}
   0 & -c & -d \\
   -c & 0 & 0 \\
   -d & 0 & 0 \\
\end{pmatrix}.
$$

For the non-symmorphic space group for the bulk $2H$ polytype, the Raman tensors are:\cite{bilbaoramantensors}

$$
\Gamma_{1}^+(A_{1g}):\begin{pmatrix}
   a & 0 & 0 \\
   0 & a & 0 \\
   0 & 0 & b \\
\end{pmatrix},
$$
$$
\Gamma_{5}^+(E_{1g}):\begin{pmatrix}
   0 & 0 & 0 \\
   0 & 0 & c \\
   0 & c & 0 \\
\end{pmatrix},
\quad
\begin{pmatrix}
   0 & 0 & -c \\
   0 & 0 & 0 \\
   -c & 0 & 0 \\
\end{pmatrix}
$$

$$
\Gamma_{6}^+(E_{2g}):\begin{pmatrix}
   d & 0 & 0 \\
   0 & -d & 0 \\
   0 & 0 & 0 \\
\end{pmatrix},
\quad
\begin{pmatrix}
   0 & -d & 0 \\
   -d & 0 & 0 \\
   0 & 0 & 0 \\
\end{pmatrix}.
$$

\section{\label{sec:level1summarydiscuss}Summary and discussions}
In this work, symmetry-related aspects of bulk and $N$-layer $2Ha$, $2Hc$ and $1T$ TMDCs polytypes were discussed from a group theory perspective. The analysis of the presence of inversion symmetry gives different behaviors (in the case of odd number of TLs) for the same number of layers in a given material, with different polytypes. Therefore, it is possible to design experiments to probe, for example, the presence of different polytypes within the same sample, with the same number of layers. The breaking of inversion symmetry is crucial in materials suitable for specific applications, like the development of valleytronic devices, and group theory predictions give directions to researches on how to design their devices to achieve their desired symmetry-related goals.

Some perturbations can lower the symmetry of these thin films and this approach has been used to tune some characteristics of these materials. In strained MoS$_2$ monolayer, where the doubly degenerate Raman active mode $E'$ splits into $E'^-$ and $E'^+$ peaks (depending on the magnitude and symmetry of the strain), an optical band gap was found and it is approximately linear with strain for both monolayer and bilayer MoS$_2$.\cite{conley2013bandgap,wang2013raman,castellanos2013local} By using different TMDCs, it is possible to engineer the optical band gap of interest to the researcher. Another possibility is the piling of different TMDCs to engineer new heterostructures, where the inversion symmetry is broken with more options made available by using multiple materials. Such heterostructures are expected, for example, to give rise to tunable band gaps from $0.79$ to $1.16$ eV.\cite{terrones2013novel}

In the present work, the symmetry properties of the vibrational modes were found for the high symmetry points and lines in the BZ, extending previous knowledge beyond the zone center phonons in TMDCs. One important aspect of this symmetry analysis is that, from symmetry variations, it is possible to predict the difference in phonon modes in these structures. $N$ new Raman-active modes have been observed in few layers TMDCs like in WSe$_{2}$.\cite{luo2013effects} Density functional theory (DFT) combined with polarization dependent Raman measurements and group theory were used to understand the first-order Raman spectra. For example, the appearance of the inactive mode $B_{2g}^{1}$ in bulk WSe$_{2}$ and only at specific laser lines is still not well understood and is usually attributed to resonance effects.\cite{luo2013effects} However, for $N$ even and $N$ odd few layers, $A_{1g}$ (for $N$ even TLs) and $A'_1$ (for $N$ odd TLs) are both observed at $310$\:cm$^{-1}$. Furthermore, the $E_{1g}$ mode at around $175$\:cm$^{-1}$ in bulk WSe$_{2}$ ($2Hc$ polytype) is not measurable under the backscattering configuration along the $z$ direction of light propagation, as well as the $E''$ mode for $1$TL of the same polytype (see the Raman tensors in section \ref{sec:level2tensors}). In films with $N\geq2$, the $E''$ mode develops into $E_{g}$ symmetry, for $N$-even TLs, and into $E'$ modes for $N$-odd layers, which are both detectable under $\overline{z}(xx)z$ and $\overline{z}(xy)z$ polarizations (and these different behaviors are not related to substrate effects, since these modes are also detected in suspended samples).\cite{luo2013effects} The mode at $260$\:cm$^{-1}$ in bulk was previously attributed to the Raman active out-of-plane $A_{1g}$ mode, but polarization measurements have shown that even for $\overline{z}(xy)z$ polarization this mode is observed, in contrast with the group theoretical prediction and the previous symmetry assignment. This mode was consequently attributed to second-order Raman scattering.\cite{luo2013effects} Similar results were observed for MoTe$_{2}$\cite{yamamoto2014strong} and are expected for other TMDCs. The complete group theory analysis described here should be used to guide researchers in making correct mode assignments using the tables and discussion given in the present work.

{\bf Acknowledgments}
The authors acknowledge financial support from CNPq grant 551953/2011-0 and NSF grant DMR-$1004147$. L.G.C. and A.J. acknowledge support from FAPEMIG.

\bibliography{bib}
\addcontentsline{toc}{section}{Referências}
\clearpage

\renewcommand{\thetable}{S \Roman{table}}

\subsection*{\label{sec:level11} Supplementary Material to ``Group Theory analysis of two-dimensional Transition Metal Dichalcogenides''}

 %\author{J. Ribeiro-Soares$^{1,2}$, R. M. Almeida$^{1}$, E. B. Barros$^{2,3}$, P. T. Araujo$^{4}$, M. S. Dresselhaus$^{2,5}$, L. G. Cançado$^{1}$ and A. Jorio$^{1}$}

J. Ribeiro-Soares$^{1,2,*}$, R. M. Almeida$^{1}$, E. B. Barros$^{2,3}$, P. T. Araujo$^{4}$, M. S. Dresselhaus$^{2,5}$, L. G. Cançado$^{1}$ and A. Jorio$^{1}$
%\footnotetex{$^*$Author to whom correspondence should be addressed: jenainassoares2@gmail.com}

$^{1}$\textit{Departamento de F\'{\i}sica, Universidade Federal de Minas Gerais, Belo Horizonte, MG, 30123-970, Brazil}

$^{2}$\textit{Department of Electrical Engineering and Computer Science, Massachusetts Institute of Technology (MIT), Cambridge, MA 02139, USA}

$^{3}$\textit{Departamento de Física, Universidade Federal do Ceará, Fortaleza, CE, 60455-900, Brazil}

$^{4}$\textit{Department of Physics and Astronomy, University of Alabama, Tuscaloosa, Alabama 35487, USA}

$^{5}$\textit{Department of Physics, Massachusetts Institute of Technology (MIT), Cambridge, MA 02139, USA}

*Author to whom correspondence should be addressed: jenainassoares2@gmail.com
\clearpage

\setcounter{table}{0}

\subsection*{\label{sec:level1} Contents}

 \renewcommand{\theenumi}{\Roman{enumi}}
 \begin{enumerate}
   \item Lattice vibration representations for bulk $2Ha$, $2Hc$ and $1T$
   \item Character tables of spacial groups modified to the group of wave vector (GWV) of each point and line of high symmetry in the BZ.
   \end{enumerate}

  \renewcommand{\theenumi}{\arabic{enumi}}
       \begin{enumerate}
         \item Spacial groups used for bulk of the $2H$ polytype
        \item Spacial groups used for $N$-odd few layers of the $2H$ polytype
        \item Spacial groups used for $N$-even few layers of the $2H$ polytype and for $N$ layer and bulk $1T$ polytype
       \end{enumerate}

\clearpage

\subsection*{I. Lattice vibration representations for bulk $2Ha$, $2Hc$ and $1T$}

In this appendix we list the lattice vibration irreducible representations $\Gamma^{vib}$ (discussed in section II~D of the main manuscript) for each high-symmetry point and line in the BZ for the bulk $2Ha$, $2Hc$ and $1T$ polytypes in Tables \ref{tab:Gamma2Habulk}, \ref{tab:Gamma2Hbbulk} and \ref{tab:Gamma1Tbulk}, respectively. The character tables of spacial groups modified to the GWV of each high-symmetry point and line of BZ are given with respect to the points and lines indicated in red in Fig. \ref{BZ} of the main manuscript.

   \begin{table*}[h!]
\caption{\label{tab:Gamma2Habulk}Wave-vector point-group representations ($\Gamma^{vib}$) for the bulk of $2Ha$-polytype (/AbA CbC/) TMDCs for all the high-symmetry points and lines in the BZ.}
%\begin{ruledtabular}
\begin{tabular}{cc}
\hline\hline
\multicolumn{2}{c}{  $2Ha$-polytype (/AbA CbC/)}\\ \hline
BZ point & Irreducible representation\\
\hline
$\Gamma$&$\Gamma_{1}^+\oplus2\Gamma_{3}^+\oplus\Gamma_{5}^+\oplus2\Gamma_{6}^+\oplus2\Gamma_{2}^-\oplus\Gamma_{4}^-\oplus2\Gamma_{5}^-\oplus\Gamma_{6}^-$ \\
 $K$&$K_{1}^+\oplus K_{2}^+\oplus4K_{3}^+\oplus2K_{1}^-\oplus2K_{2}^-\oplus2K_{3}^-$ \\
 $M$&$3M_{1}^+\oplus2M_{2}^+\oplus3M_{3}^+\oplus M_{4}^+\oplus M_{1}^-\oplus3M_{2}^-\oplus2M_{3}^-\oplus3M_{4}^-$ \\
 $\Sigma$&$6\Sigma_{1}\oplus2\Sigma_{2}\oplus4\Sigma_{3}\oplus6\Sigma_{4}$ \\
 $T (T')$&$5T_{1}\oplus4T_{2}\oplus5T_{3}\oplus4T_{4}$ \\
 $u$&$10u^+\oplus8u^-$ \\
 \hline\hline
\end{tabular}
%\end{ruledtabular}
    \end{table*}

   \begin{table*}[h!]
\caption{\label{tab:Gamma2Hbbulk}Wave-vector point-group representations ($\Gamma^{vib}$) for the bulk of $2Hc$-polytype (/CaC AcA/) TMDCs for all the high-symmetry points and lines in the BZ.}
\begin{tabular}{cc}
\hline\hline
\multicolumn{2}{c}{  $2Hc$-polytype (/CaC AcA/)} \\ \hline
BZ point & Irreducible representation\\
\hline
$\Gamma$&$\Gamma_{1}^+\oplus2\Gamma_{3}^+\oplus\Gamma_{5}^+\oplus2\Gamma_{6}^+\oplus2\Gamma_{2}^-\oplus\Gamma_{4}^-\oplus2\Gamma_{5}^-\oplus\Gamma_{6}^-$ \\
 $K$&$2K_{1}^+\oplus2K_{2}^+\oplus3K_{3}^+\oplus K_{1}^-\oplus K_{2}^-\oplus3K_{3}^-$ \\
 $M$&$3M_{1}^+\oplus2M_{2}^+\oplus3M_{3}^+\oplus M_{4}^+\oplus M_{1}^-\oplus3M_{2}^-\oplus2M_{3}^-\oplus3M_{4}^-$ \\
 $\Sigma$&$6\Sigma_{1}\oplus2\Sigma_{2}\oplus4\Sigma_{3}\oplus6\Sigma_{4}$ \\
 $T (T')$&$5T_{1}\oplus4T_{2}\oplus5T_{3}\oplus4T_{4}$ \\
 $u$&$10u^+\oplus8u^-$ \\
 \hline\hline
\end{tabular}
    \end{table*}

 \begin{table*}[h!]
\caption{\label{tab:Gamma1Tbulk}Wave-vector point-group representations ($\Gamma^{vib}$) for the bulk of $1T$-polytype (/AbC/AbC/) TMDCs for all the high-symmetry points and lines in the BZ.}
\begin{tabular}{cc}
\hline\hline
\multicolumn{2}{c}{  $1T$-polytype (/AbC/AbC/) } \\ \hline
BZ point & Irreducible representation\\
\hline
$\Gamma$&$\Gamma_{1}^+\oplus \Gamma_{3}^+\oplus2\Gamma_{2}^-\oplus2\Gamma_{3}^-$ \\
 $K$&$K_{1}\oplus2K_{2}\oplus3K_{3}$ \\
 $M$&$2M_{1}^+\oplus 2M_{1}^-\oplus M_{2}^+\oplus4M_{2}^-$ \\
 $\Sigma$&$6\Sigma_{1}\oplus3\Sigma_{2}$ \\
 $T (T')$&$4T_{1}\oplus5T_{2}$ \\
 $u$&$9u$ \\
 \hline\hline
\end{tabular}
    \end{table*}

\subsection*{II. Character tables of spacial groups modified to the group of the wave vector (GWV) of each point and line of high symmetry in the BZ.}

Tables \ref{tab:D6hbulk} to \ref{tab:C1hubulk} give the character tables for the GWV for the $2Ha$ and $2Hc$ bulk polytypes. Tables \ref{tab:D3hodd} to \ref{tab:Cxyuodd} give the character tables to the GWV for the $2H$ polytype with $N$-odd layers, while Tables \ref{tab:D3deven} to \ref{tab:Cxzssigmaeven} give the character tables for the GWV for the $2H$ polytype with $N$-even layers. The space group for the $1T$ bulk polytype, as well as that for $N$-even and $N$-odd layers (the $1T$ bulk polytype is symmorphic) is the $P\bar{3}m1$ ($D^{3}_{3d}$ or \#$164$) and the GWV for each high-symmetry point or line in the BZ is the same, regardless of the number of layers. The GWV for each high-symmetry point or line in the BZ for the $1T$ polytype is the same as that which occurs in the $2H$ polytype with an even number of layers, and the Tables \ref{tab:D3deven} to \ref{tab:Cxzssigmaeven} should be used for this polytype. The tables contain the Space Group (SG) and Point Group (PG) notation for the irreducible representations, and they are given in the following order:

\renewcommand{\theenumi}{\arabic{enumi}}
       \begin{enumerate}
        \item Spacial groups used for bulk of the $2H$ polytype
        \item Spacial groups used for $N$-odd few layers of the $2H$ polytype
        \item Spacial groups used for $N$-even few layers of the $2H$ polytype and for $N$ layer and bulk $1T$ polytype.
       \end{enumerate}

\clearpage

\subsection*{1. Spacial groups used for bulk of the $2H$ polytype}

\begin{table*}[h!]
\caption{\label{tab:D6hbulk} Character table for the $\Gamma$ point [$D^{4}_{6h}$ ($P6_3/mmc$, \#194)].}
\vspace{0.3cm}\tiny
\begin{ruledtabular}
\begin{tabular}{ccccccccccccccc}

 &  &  &  & &  & $\{C'^{A}_{2}|0\}$ & $\{C''^{A}_{2}|0\}$ &  &  &  &  & $\{\sigma^{A}_{d}|0\}$ & $\{\sigma^{A}_{v}|\tau\}$\footnote{$\tau$ is the translation of half of the $c$ lattice parameter along the $\hat{z}$ direction ($\tau=(\frac{1}{2})c\hat{z}$).\label{fn:repeattau}}\\
&  & & $\{C^{+}_{3}|0\}$ &  & $\{C^{-}_{6}|\tau\}$\footref{fn:repeattau} & $\{C'^{B}_{2}|0\}$ & $\{C''^{B}_{2}|0\}$ &  & $\{S^{+}_{6}|0\}$ &  & $\{S^{-}_{3}|0\}$ & $\{\sigma^{B}_{d}|0\}$ & $\{\sigma^{B}_{v}|\tau\}$ \\
SG & PG & $\{E|0\}$ & $\{C^{-}_{3}|0\}$ & $\{C_{2}|\tau\}$\footref{fn:repeattau} & $\{C^{+}_{6}|\tau\}$ & $\{C'^{C}_{2}|0\}$ & $\{C''^{C}_{2}|0\}$ & $\{i|0\}$ &  $\{S^{-}_{6}|0\}$ & $\{\sigma_{h}|0\}$ & $\{S^{+}_{3}|0\}$ & $\{\sigma^{C}_{d}|0\}$ & $\{\sigma^{C}_{v}|\tau\}$ & Bases \\
\hline
 $\Gamma^{+}_{1}$ & $A_{1g}$ & $1$ & $1$ & $1$ & $1$ & $1$ & $1$ & $1$ & $1$ & $1$ & $1$ & $1$ & $1$ & $x^{2}+y^{2},z^{2}$ \\
 $\Gamma^{+}_{2}$ & $A_{2g}$ & $1$ & $1$ & $1$ & $1$ & $-1$ & $-1$ & $1$ & $1$ & $1$ & $1$ & $-1$ & $-1$ &  \\
 $\Gamma^{+}_{3}$ & $B_{1g}$ & $1$ & $1$ & $-1$ & $-1$ & $-1$ & $1$ & $1$ & $1$ & $-1$ & $-1$ & $-1$ & $1$ & \\
 $\Gamma^{+}_{4}$ & $B_{2g}$ & $1$ & $1$ & $-1$ & $-1$ & $1$ & $-1$ & $1$ & $1$ & $-1$ & $-1$ & $1$ & $-1$ & \\
 $\Gamma^{+}_{5}$ & $E_{1g}$ & $2$ & $-1$ & $-2$ & $1$ & $0$ & $0$ & $2$ & $-1$ & $-2$ & $1$ & $0$ & $0$ & $ (xz, yz)$ \\
 $\Gamma^{+}_{6}$ & $E_{2g}$ & $2$ & $-1$ & $2$ & $-1$ & $0$ & $0$ & $2$ & $-1$ & $2$ & $-1$ & $0$ & $0$ & $(xy,x^{2}-y^{2})$ \\
 $\Gamma^{-}_{1}$ & $A_{1u}$ & $1$ & $1$ & $1$ & $1$ & $1$ & $1$ & $-1$ & $-1$ & $-1$ & $-1$ & $-1$ & $-1$ & \\
 $\Gamma^{-}_{2}$ & $A_{2u}$ & $1$ & $1$ & $1$ & $1$ & $-1$ & $-1$ & $-1$ & $-1$ & $-1$ & $-1$ & $1$ & $1$ & $z$ \\
 $\Gamma^{-}_{3}$ & $B_{1u}$ & $1$ & $1$ & $-1$ & $-1$ & $-1$ & $1$ & $-1$ & $-1$ & $1$ & $1$ & $1$ & $-1$ & \\
 $\Gamma^{-}_{4}$ & $B_{2u}$ & $1$ & $1$ & $-1$ & $-1$ & $1$ & $-1$ & $-1$ & $-1$ & $1$ & $1$ & $-1$ & $1$ & \\
 $\Gamma^{-}_{5}$ & $E_{1u}$ & $2$ & $-1$ & $-2$ & $1$ & $0$ & $0$ & $-2$ & $1$ & $2$ & $-1$ & $0$ & $0$ & $(x,y)$ \\
 $\Gamma^{-}_{6}$ & $E_{2u}$ & $2$ & $-1$ & $2$ & $-1$ & $0$ & $0$ & $-2$ & $1$ & $-2$ & $1$ & $0$ & $0$ & \\
\end{tabular}
\end{ruledtabular}
    \end{table*}
%%%%%%%%%%%%%%%%%%%%%%%%%%%%%%%%%%%%%%%%%%%%%%%%%%%%%%%%%%%%%%%%%%%%%

 \begin{table*}[h!]
\caption{\label{tab:D3hbulk} Character table for the $K$($K'$) point [$D^{4}_{3h}$ ($P\bar{6}2c$, \#190)].}
\begin{ruledtabular}
\begin{tabular}{ccccccccc}

 &  &  &  & $\{C'^{A}_{2}|0\}$ &  &  & $\{\sigma^{A}_{v}|\tau\}$\footnote{$\tau$ is the translation of half of the $c$ lattice parameter along the $\hat{z}$ direction ($\tau=(\frac{1}{2})c\hat{z}$).\label{fn:repeattau}} \\
 &  &  & $\{C^{+}_{3}|0\}$ & $\{C'^{B}_{2}|0\}$ &  & $\{S^{-}_{3}|0\}$ & $\{\sigma^{B}_{v}|\tau\}$\\
 SG & PG & $\{E|0\}$ & $\{C^{-}_{3}|0\}$ & $\{C'^{C}_{2}|0\}$ & $\{\sigma_{h}|0\}$ & $\{S^{+}_{3}|0\}$ & $\{\sigma^{C}_{v}|\tau\}$ & Bases \\
\hline
 $K^{+}_{1}$ & $A^{\prime}_1$ & $1$ & $1$ & $1$ & $1$ & $1$ & $1$ & $x^{2}+y^{2},z^{2}$ \\
 $K^{+}_{2}$ & $A^{\prime}_2$ & $1$ & $1$ & $-1$ & $1$ & $1$ & $-1$ &  \\
 $K^{+}_{3}$ & $E^{\prime}$ & $2$ & $-1$ & $0$ & $2$ & $-1$ & $0$ & $(x,y), (xy, x^{2}-y^{2})$ \\
 $K^{-}_{1}$ & $A^{\prime\prime}_1$ & $1$ & $1$ & $1$ & $-1$ & $-1$ & $-1$ & \\
 $K^{-}_{2}$ & $A^{\prime\prime}_2$ & $1$ & $1$ & $-1$ & $-1$ & $-1$ & $1$ & $z$ \\
 $K^{-}_{3}$ & $E^{\prime\prime}$ & $2$ & $-1$ & $0$ & $-2$ & $1$ & $0$ & $ (yz, xz)$ \\
\end{tabular}
\end{ruledtabular}
    \end{table*}
%%%%%%%%%%%%%%%%%%%%%%%%%%%%%%%%%%%%%%%%%%%%%%%%%%%%%%%%%%%%%%%%%%%%%

\begin{table*}[h!]
\caption{\label{tab:D2hbulk} Character table for the $M$ point [$D^{17}_{2h}$ ($Cmcm$, \#63)].}
\begin{ruledtabular}
\begin{tabular}{ccccccccccc}

  SG & PG & $\{E|0\}$ & $\{C_{2}|\tau\}$\footnote{$\tau$ is the translation of half of the $c$ lattice parameter along the $\hat{z}$ direction ($\tau=(\frac{1}{2})c\hat{z}$).\label{fn:repeattau}} & $\{C'^{A}_{2}|0\}$ & $\{C''^{A}_{2}|0\}$ & $\{i|0\}$ & $\{\sigma_{h}|0\}$ & $\{\sigma^{A}_{d}|0\}$ & $\{\sigma^{A}_{v}|\tau\}$\footref{fn:repeattau} & Bases \\
\hline
 $M^{+}_{1}$ & $A_{g}$ & $1$ & $1$ & $1$ & $1$ & $1$ & $1$ & $1$ & $1$ & $x^{2}, y^{2}, z^{2}$ \\
 $M^{+}_{2}$ & $B_{1g}$ & $1$ & $1$ & $-1$ & $-1$ & $1$ & $1$ & $-1$ & $-1$ & $ xy$ \\
 $M^{+}_{3}$ & $B_{2g}$ & $1$ & $-1$ & $1$ & $-1$ & $1$ & $-1$ & $1$ & $-1$ & $ xz$ \\
 $M^{+}_{4}$ & $B_{3g}$ & $1$ & $-1$ & $-1$ & $1$ & $1$ & $-1$ & $-1$ & $1$ & $ yz$ \\
 $M^{-}_{1}$ & $A_{u}$ & $1$ & $1$ & $1$ & $1$ & $-1$ & $-1$ & $-1$ & $-1$ & \\
 $M^{-}_{2}$ & $B_{1u}$ & $1$ & $1$ & $-1$ & $-1$ & $-1$ & $-1$ & $1$ & $1$ & $z$ \\
 $M^{-}_{3}$ & $B_{2u}$ & $1$ & $-1$ & $1$ & $-1$ & $-1$ & $1$ & $-1$ & $1$ & $y$ \\
 $M^{-}_{4}$ & $B_{3u}$ & $1$ & $-1$ & $-1$ & $1$ & $-1$ & $1$ & $1$ & $-1$ & $x$ \\
\end{tabular}
\end{ruledtabular}
    \end{table*}

%%%%%%%%%%%%%%%%%%%%%%%%%%%%%%%%%%%%%%%%%%%%%%%%%%%%%%%%%%%%%%%%%%%%%

    \begin{table*}[h!]
\caption{\label{tab:C2vTbulk} Character table for the $T$($T'$) line [$C^{16}_{2v}$ ($Ama2$, \#40)].}
\begin{ruledtabular}
\begin{tabular}{ccccccc}
  SG & PG & $\{E|0\}$ & $\{C'^{A}_{2}|0\}$ & $\{\sigma_{h}|0\}$ & $\{\sigma^{A}_{v}|\tau\}$\footnote{$\tau$ is the translation of half of the $c$ lattice parameter along the $\hat{z}$ direction ($\tau=(\frac{1}{2})c\hat{z}$).\label{fn:repeattau}} & Bases \\
\hline
 $T_{1}$ & $A_{1}$ & $1$ & $1$ & $1$ & $1$ & $y, x^{2}, y^{2}, z^{2}$ \\
 $T_{2}$ & $A_{2}$ & $1$ & $1$ & $-1$ & $-1$ & $ xz $ \\
 $T_{3}$ & $B_{1}$ & $1$ & $-1$ & $1$ & $-1$ & $x, xy$ \\
 $T_{4}$ & $B_{2}$ & $1$ & $-1$ & $-1$ & $1$ & $z, yz$ \\
\end{tabular}
\end{ruledtabular}
    \end{table*}
%%%%%%%%%%%%%%%%%%%%%%%%%%%%%%%%%%%%%%%%%%%%%%%%%%%%%%%%%%%%%%%%%%%%%

    \begin{table*}[h!]
\caption{\label{tab:C2vsigmabulk} Character table for the $\Sigma$ line [$C^{14}_{2v}$ ($Amm2$, \#38)].}
\begin{ruledtabular}
\begin{tabular}{ccccccc}
 SG & PG & $\{E|0\}$ & $\{C''^{A}_{2}|0\}$ & $\{\sigma_{h}|0\}$ & $\{\sigma^{A}_{d}|0\}$ & Bases \\
\hline
$\Sigma_{1}$ & $A_{1}$ & $1$ & $1$ & $1$ & $1$ & $x, x^{2}, y^{2}, z^{2}$ \\
$\Sigma_{2}$ & $A_{2}$ & $1$ & $1$ & $-1$ & $-1$ & $ yz $ \\
$\Sigma_{3}$ & $B_{1}$ & $1$ & $-1$ & $1$ & $-1$ & $y, xy$ \\
$\Sigma_{4}$ & $B_{2}$ & $1$ & $-1$ & $-1$ & $1$ & $z, xz$ \\
\end{tabular}
\end{ruledtabular}
    \end{table*}
%%%%%%%%%%%%%%%%%%%%%%%%%%%%%%%%%%%%%%%%%%%%%%%%%%%%%%%%%%%%%%%%%%%%%
    \begin{table*}[h!]
\caption{\label{tab:C1hubulk} Character table for the $u$ point [$C^{xy}_{s}$ or $C^{1}_{s}$, $Pm$, \#6]. The $\sigma_{h}$ mirror plane lies in the xy plane.}
\begin{ruledtabular}
\begin{tabular}{ccccc}
 SG & PG & $\{E|0\}$ & $\{\sigma_{h}|0\}$ & Bases \\
\hline
 $u^{+}$ & $A^{\prime}$ & $1$ & $1$ & $x, y, x^{2}, y^{2}, z^{2}, xy$ \\
 $u^{-}$ & $A^{\prime\prime}$ & $1$ & $-1$ & $z, yz, xz$ \\
\end{tabular}
\end{ruledtabular}
    \end{table*}

\clearpage
%%%%%%%%%%%%%%%%%%%%%%%%%%%%%%%%%%%%%%%%%%%%%%%%%%%%%%%%%%%%%%%%%%%%%

\subsection*{2. Spacial groups used for $N$-odd few layers of the $2H$ polytype}

 \begin{table*}[h!]
\caption{\label{tab:D3hodd} Character table for the $\Gamma$ point [$D^{1}_{3h}$ ($P\bar{6}m2$, \#187)].}
\begin{ruledtabular}
\begin{tabular}{ccccccccc}
  &  &  &  & $C'^A_{2}$ &  &  & $\sigma^A_{v}$ &  \\
  &  &  & $C^+_{3}$ & $C'^B_{2}$ &  & $S^-_{3}$ & $\sigma^B_{v}$ &  \\
 SG & PG & $E$ & $C^-_{3}$ & $C'^C_{2}$ & $\sigma_{h}$ & $S^+_{3}$ & $\sigma^C_{v}$ & Bases \\
\hline
 $\Gamma^{+}_{1}$& $A^{\prime}_1$ & $1$ & $1$ & $1$ & $1$ & $1$ & $1$ & $x^{2}+y^{2},z^{2}$ \\
 $\Gamma^{+}_{2}$& $A^{\prime}_2$ & $1$ & $1$ & $-1$ & $1$ & $1$ & $-1$ &  \\
 $\Gamma^{+}_{3}$& $E^{\prime}$ & $2$ & $-1$ & $0$ & $2$ & $-1$ & $0$ & $(x,y), (xy, x^{2}-y^{2})$ \\
 $\Gamma^{-}_{1}$& $A^{\prime\prime}_1$ & $1$ & $1$ & $1$ & $-1$ & $-1$ & $-1$ & \\
 $\Gamma^{-}_{2}$& $A^{\prime\prime}_2$ & $1$ & $1$ & $-1$ & $-1$ & $-1$ & $1$ & $z$ \\
 $\Gamma^{-}_{3}$& $E^{\prime\prime}$ & $2$ & $-1$ & $0$ & $-2$ & $1$ & $0$ & $(yz, xz)$ \\
\end{tabular}
\end{ruledtabular}
    \end{table*}
%%%%%%%%%%%%%%%%%%%%%%%%%%%%%%%%%%%%%%%%%%%%%%%%%%%%%%%%%%%%%%%%%%%%%

\begin{table*}[h!]
\caption{\label{tab:C3hodd} Character table for the $K$($K'$) point [$C^{1}_{3h}$ ($P\bar{6}$, \#174)].}
\begin{ruledtabular}
\begin{tabular}{ccccccccccc}
 SG & PG & $E$ & $C^{+}_{3}$ & $C^{-}_{3}$ & $\sigma_{h}$ & $S^{+}_{3}$ & $S^{-}_{3}$ & Bases \\
\hline
 $K^+_1$ & $A^{\prime}$ & $1$ & $1$ & $1$ & $1$ & $1$ & $1$ & $x^{2}+y^{2}, z^{2}$ \\
 $K^-_1$ & $A^{\prime\prime}$ & $1$ & $1$ & $1$ & $-1$ & $-1$ & $-1$ & $z$ \\
 $K^+_2$ & $E^{\prime}$ & $1$ & $\omega$\footnote{$\omega$ $=$ $\exp{(2i\pi/3)}$} & $\omega^{2}$ & $1$ & $\omega$ & $\omega^{2}$ & \multirow{2}{*}{\Bigg\}$(x, y), (x^{2}-y^{2}, xy)$} \\
 $K^{+*}_2$ & $E^{\prime*}$ & $1$ & $\omega^{2}$ & $\omega$ & $1$ & $\omega^{2}$ & $\omega$ & \\
 $K^-_2$ & $E^{\prime\prime}$ & $1$ & $\omega$ & $\omega^{2}$ & $-1$ & $-\omega$ & $-\omega^{2}$ & \multirow{2}{*}{\Bigg\}$(xz, yz)$}\\
 $K^{-*}_2$ & $E^{\prime\prime*}$ & $1$ & $\omega^{2}$ & $\omega$ & $-1$ & $-\omega^{2}$ & $-\omega$ &  \\
\end{tabular}
\end{ruledtabular}
    \end{table*}

%%%%%%%%%%%%%%%%%%%%%%%%%%%%%%%%%%%%%%%%%%%%%%%%%%%%%%%%%%%%%%%%%%%%%

    \begin{table*}[h!]
\caption{\label{tab:C2voddM} Character for the $M$ point [$C^{14}_{2v}$ ($Amm2$, \#38)].}
\begin{ruledtabular}
\begin{tabular}{ccccccc}
  SG & PG & $E$ & $C'^{A}_{2}$ & $\sigma_{h}$ & $\sigma^{A}_{v}$ & Bases \\
\hline
 $M_{1}$ & $A_1$ & $1$ & $1$ & $1$ & $1$ & $x, x^{2}, y^{2}, z^{2}$ \\
 $M_{2}$ & $A_2$ & $1$ & $1$ & $-1$ & $-1$ & $ yz $ \\
 $M_{3}$ & $B_1$ & $1$ & $-1$ & $1$ & $-1$ & $y, xy$ \\
 $M_{4}$ & $B_2$ & $1$ & $-1$ & $-1$ & $1$ & $z, xz$ \\
\end{tabular}
\end{ruledtabular}
    \end{table*}
%%%%%%%%%%%%%%%%%%%%%%%%%%%%%%%%%%%%%%%%%%%%%%%%%%%%%%%%%%%%%%%%%%%%%

    \begin{table*}[h!]
\caption{\label{tab:CxyTodd} Character table for the $T$($T'$) line [$C^{xy}_{s}$ or $C^{1}_{s}$, $Pm$, \#6]. The $\sigma_{h}$ mirror plane lies on xy plane.}
\begin{ruledtabular}
\begin{tabular}{ccccc}
 SG & PG & $E$ & $\sigma_{h}$ & Bases \\
\hline
 $T^+$ & $A^{\prime}$ & $1$ & $1$ & $x, y, x^{2}, y^{2}, z^{2}, xy$ \\
 $T^-$ & $A^{\prime\prime}$ & $1$ & $-1$ & $z, yz, xz$ \\
\end{tabular}
\end{ruledtabular}
    \end{table*}
%%%%%%%%%%%%%%%%%%%%%%%%%%%%%%%%%%%%%%%%%%%%%%%%%%%%%%%%%%%%%%%%%%%%%

 \begin{table*}[h!]
\caption{\label{tab:C2vsigmaodd} Character table for the $\Sigma$ line [$C^{14}_{2v}$ ($Amm2$, \#38)].}
\begin{ruledtabular}
\begin{tabular}{ccccccc}
  SG & PG & $E$ & $C'^{A}_{2}$ & $\sigma_{h}$ & $\sigma^{A}_{v}$ & Bases \\
\hline
$\Sigma_{1}$ & $A_1$ & $1$ & $1$ & $1$ & $1$ & $x, x^{2}, y^{2}, z^{2}$ \\
$\Sigma_{2}$ & $A_2$ & $1$ & $1$ & $-1$ & $-1$ & $ yz $ \\
$\Sigma_{3}$ & $B_1$ & $1$ & $-1$ & $1$ & $-1$ & $y, xy$ \\
$\Sigma_{4}$ & $B_2$ & $1$ & $-1$ & $-1$ & $1$ & $z, xz$ \\
\end{tabular}
\end{ruledtabular}
    \end{table*}
%%%%%%%%%%%%%%%%%%%%%%%%%%%%%%%%%%%%%%%%%%%%%%%%%%%%%%%%%%%%%%%%%%%%%

    \begin{table*}[h!]
\caption{\label{tab:Cxyuodd} Character table for the $u$ point [$C^{xy}_{s}$ or $C^{1}_{s}$, $Pm$, \#6]. The $\sigma_{h}$ mirror lies on $xy$ plane.}
\begin{ruledtabular}
\begin{tabular}{ccrrc}
 SG & PG & $E$ & $\sigma_{h}$ & Bases \\
\hline
 $u^+$ & $A^{\prime}$ & $1$ & $1$ & $x, y, x^{2}, y^{2}, z^{2}, xy$ \\
 $u^-$ & $A^{\prime\prime}$ & $1$ & $-1$ & $z, yz, xz$ \\
\end{tabular}
\end{ruledtabular}
    \end{table*}
%%%%%%%%%%%%%%%%%%%%%%%%%%%%%%%%%%%%%%%%%%%%%%%%%%%%%%%%%%%%%%%%%%%%%

\clearpage
\subsection*{3. Spacial groups used for $N$-even few layers of the $2H$ polytype and for $N$ layer and bulk $1T$ polytype}

%%%%%%%%%%%%%%%%%%%%%%%%%%%%%%%%%%%%%%%%%%%%%%%%%%%%%%%%%%%%%%%%%%%%%
\begin{table*}[h!]
\caption{\label{tab:D3deven} Character table for the $\Gamma$ point [$D^{3}_{3d}$ ($P\bar{3}m1$, \#164)].}
\begin{ruledtabular}
\begin{tabular}{ccccccccc}
  &  &  &  & $C'^A_{2}$ &  &  & $\sigma^A_{d}$ &  \\
  &  &  & $C^+_{3}$ & $C'^B_{2}$ &  & $S^+_{6}$ & $\sigma^B_{d}$ &  \\
 SG & PG & $E$ & $C^-_{3}$ & $C'^C_{2}$ & $i$ & $S^-_{6}$ & $\sigma^C_{d}$ & Bases \\
\hline
$\Gamma^{+}_{1}$ & $A_{1g}$ & $1$ & $1$ & $1$ & $1$ & $1$ & $1$ & $x^{2}+y^{2}, z^{2}$ \\
$\Gamma^{+}_{2}$ & $A_{2g}$ & $1$ & $1$ & $-1$ & $1$ & $1$ & $-1$ &  \\
$\Gamma^{+}_{3}$ & $E_{g}$ & $2$ & $-1$ & $0$ & $2$ & $-1$ & $0$ & $(xz, yz), (x^{2}-y^{2}, xy)$ \\
$\Gamma^{-}_{1}$ & $A_{1u}$ & $1$ & $1$ & $1$ & $-1$ & $-1$ & $-1$ &  \\
$\Gamma^{-}_{2}$ & $A_{2u}$ & $1$ & $1$ & $-1$ & $-1$ & $-1$ & $1$ & $z$ \\
$\Gamma^{-}_{3}$ & $E_{u}$ & $2$ & $-1$ & $0$ & $-2$ & $1$ & $0$ & $(x, y)$ \\
\end{tabular}
\end{ruledtabular}
    \end{table*}
%%%%%%%%%%%%%%%%%%%%%%%%%%%%%%%%%%%%%%%%%%%%%%%%%%%%%%%%%%%%%%%%%%%%%

\begin{table*}[h!]
\caption{\label{tab:D3even} Character table for the $K$($K'$) point [$D^{2}_{3}$ ($P321$, \#150)].}
\begin{ruledtabular}
\begin{tabular}{cccccc}
  &  &  &  & $C'^A_{2}$ &  \\
  &  &  & $C^+_{3}$ & $C'^B_{2}$ &  \\
 SG & PG & $E$ & $C^-_{3}$ & $C'^C_{2}$ & Bases \\
\hline
 $K_{1}$ & $A_{1}$ & $1$ & $1$ & $1$ & $x^{2}+y^{2}, z^{2}$ \\
 $K_{2}$ & $A_{2}$ & $1$ & $1$ & $-1$ & $z$ \\
 \multirow{2}{*}{$K_{3}$} & \multirow{2}{*}{$E$} & \multirow{2}{*}{$2$} & \multirow{2}{*}{$-1$} & \multirow{2}{*}{$0$} & \multirow{2}{*}{$(xz, yz), (x, y)$} \\
  &  &  &  & & $(x^{2}-y^{2}, xy)$ \\
\end{tabular}
\end{ruledtabular}
    \end{table*}
%%%%%%%%%%%%%%%%%%%%%%%%%%%%%%%%%%%%%%%%%%%%%%%%%%%%%%%%%%%%%%%%%%%%%

\begin{table*}[h!]
\caption{\label{tab:C2hevenM} Character table for the $M$ point [$C^{3}_{2h}$ ($C2/m$, \#12)].}
\begin{ruledtabular}
\begin{tabular}{ccccccc}
 SG & PG & $E$ & $C'^{A}_{2}$ & $\sigma^{A}_{d}$ & $i$ & Bases \\
\hline
 $M^{+}_{1}$ & $A_{g}$ & $1$ & $1$ & $1$ & $1$ & $x^{2}, y^{2}, z^{2}, xz$ \\
 $M^{-}_{1}$ & $A_{u}$ & $1$ & $1$ & $-1$ & $-1$ & $y$ \\
 $M^{+}_{2}$ & $B_{g}$ & $1$ & $-1$ & $-1$ & $1$ & $xy, yz$ \\
 $M^{-}_{2}$ & $B_{u}$ & $1$ & $-1$ & $1$ & $-1$ & $x, z$ \\
\end{tabular}
\end{ruledtabular}
    \end{table*}
%%%%%%%%%%%%%%%%%%%%%%%%%%%%%%%%%%%%%%%%%%%%%%%%%%%%%%%%%%%%%%%%%%%%%

\begin{table*}[h!]
\caption{\label{tab:C2Teven} Character table for the $T$($T'$) line [$C^{3}_{2}$ ($C2$, \#5)].}
\begin{ruledtabular}
\begin{tabular}{ccccc}
 SG & PG & $E$ & $C'^{A}_{2}$ & Bases \\
\hline
 $T_1$ & A & $1$ & $1$ & $y, x^{2}, y^{2}, z^{2}, xz$ \\
 $T_2$ & B & $1$ & $-1$ & $x, z, xy, yz$ \\
\end{tabular}
\end{ruledtabular}
    \end{table*}
%%%%%%%%%%%%%%%%%%%%%%%%%%%%%%%%%%%%%%%%%%%%%%%%%%%%%%%%%%%%%%%%%%%%%

\begin{table*}[h!]
\caption{\label{tab:Cxzssigmaeven} Character table for the $\Sigma$ line [$C^{xz}_{s}$ or $C^{3}_{s}$, $Cm$, \#8]. The $\sigma$ mirror plane lies in the $xz$ plane.}
\begin{ruledtabular}
\begin{tabular}{ccccc}
 SG & PG & $E$ & $\sigma^A_d$ & Bases \\
\hline
 $\Sigma_1$ & $A^{\prime}$ & $1$ & $1$ & $x, z, x^{2}, y^{2}, z^{2}, xz$ \\
 $\Sigma_2$ & $A^{\prime\prime}$ & $1$ & $-1$ & $ y, xy, yz$ \\
\end{tabular}
\end{ruledtabular}
    \end{table*}
%%%%%%%%%%%%%%%%%%%%%%%%%%%%%%%%%%%%%%%%%%%%%%%%%%%%%%%%%%%%%%%%%%%%%

\begin{table*}[h!]
\caption{\label{tab:pgd6h} Character table for the $u$ point [$C^{1}_{1}$ ($P1$, \#1)].}
\begin{ruledtabular}
\begin{tabular}{cccc}
 SG & PG & $E$ & Bases \\
\hline
 $u$ & A & $1$ & any $f(x, y, z)$ \\
\end{tabular}
\end{ruledtabular}
    \end{table*}

\end{document}